\documentclass[11pt]{article}
\usepackage{epsf}
\usepackage{graphicx}        

\def\ll{\label}

\def\c{\cite}

\def\r1{(\ref{$1})}

\def\ba{\begin{array}{c}}

\def\ea{\end{array}}

\def\l{\left}
\def\l({\left(}
\def\r){\right)}
\def\r{\right}

\def\la{\lambda}

 \def\be{\begin{equation}}
\def\bc{\begin{center}}
\def\ec{\end{center}}
\def\bit{\begin{itemize}}
\def\eit{\end{itemize}}
\def\ee{\end{equation}}
\def\ed{\end{document}}
\def\bea{\begin{eqnarray}}
\def\eea{\end{eqnarray}}
\def\efr{\end{flushright}}

\begin{document}
\title{{ 
 Two-fold integrable hierarchy of nonholonomic deformation  
of the
DNLS
   and the Lenells-Fokas equation
}}
\author{ Anjan Kundu\\
 Theory Group \& CAMCS,
Saha Institute of Nuclear Physics\\
 Calcutta, INDIA\\
{anjan.kundu@saha.ac.in}} \maketitle
\noindent Short Title: {\it Integrable
 deformed DNLS }
\vskip 1cm

\begin{abstract}
The concept of the nonholonomic deformation formulated recently
 for the AKNS family
is extended to the Kaup-Newell class. Applying this construction we discover 
 a novel  mixed
integrable  hierarchy related to the deformed  derivative nonlinear
 Schr\"odinger (DNLS) equation and found the exact soliton
solutions exhibiting unusual accelerating motion
 for both its field and the perturbing functions. 
Extending the idea of deformation  the integrable perturbation
of the gauge related Chen-Lee-Liu DNLS 
equation is constructed together with its soliton solution. 
We  show that, the recently proposed   Lenells-Fokas (LF)
 equation   falls in 
the    deformed DNLS   
 hierarchy,    sharing the accelerating soliton   
 and  other  unusual   features. Higher order
 integrable deformations of the LF
 and the DNLS equations are  proposed.
\end{abstract}

\vskip .4cm

\noindent {\bf I. INTRODUCTION}

Intensive research  extended over    fifty years,
starting from the  pioneering   work of GGKM \c{ggkm65} 
has taken  the theory of  nonlinear integrable systems
to a well established subject, seemingly with   no  further surprises
could be expected from it. However, quite recently 
 an integrable 6th-order Korteweg de Vris (KdV) equation  was  
discovered \c{6kdv},  apparently
contradicting
the conventional wisdom.  Though eventually  this was 
 understood  as an nonholonomic
deformation  of the standard KdV equation
\c{kuper08,kundu082,chineese08},    
this   unconventional concept 
  stimulated further study  to   unveil    the
    nonholonomic deformation for the entire AKNS family \c{JMP09}.
 Consequently,  deformed 
integrable  hierarchies were discovered for the     
 nonlinear Schr\"odinger (NLS) equation,  the  sine-Gordon equation
 and
the   modified KdV 
equation \c{JMP09,kundu082, kundu083}, though the basic idea,
 exploiting the 
negative flow in the spectral parameter, was  already
 used in some specific cases \c{stramp,dmkdv09}.
Such deformations for the field theoretical models can  be interpreted also 
 as a perturbation of the
original integrable system with certain differential constraint
on the perturbing function, such that the perturbed system 
together with their deformed hierarchy as a whole  becomes a new
integrable system. 
However, in spite of  the natural expectation \c{JMP09}, the 
nonholonomic deformation could not be constructed for  the  Kaup-Newell (KN)
class  \c{KN}, which includes the important integrable system like the 
 derivative 
NLS (DNLS)  equation (\ref{dnls}).

In such a situation,
  another unexpected integrable equation (\ref{ut}) was proposed 
 very recently
by Lenells and Fokas (LF)
 \c{LF}, soliton solution of which exhibits close resemblance
with the solution of the DNLS equation. 
    This intriguing fact gave us the right hint to
 formulate  the nonholonomic deformation  for the KN spectral problem  and
applying this construction discover 
 a novel  mixed integrable hierarchy for the deformed
 DNLS family, yielding more general
  accelerating 
 soliton solutions.
This   unusual soliton dynamics is sustained by the time-dependent 
 asymptotic  of the 
perturbing functions,
 which   themselves  take a consistent   soli tonic form.
Solving the differential constraints 
 it is possible  in some cases to express 
the perturbing functions locally through the basic field,
reducing the perturbed system to a novel higher nonlinear equation in terms
of the
basic or its potential field.   

Most interestingly, we could    identify  the LF equation as the first
nontrivial member
in the mixed  hierarchy of the deformed  DNLS,
 which  enables us to  bypass  the
     lengthy and involved approaches like the Riemann-Hilbert (RH)
 or the dressing
method for extracting the soliton
solution of the LF equation  adopted in \c{LF,lenells09}
  and  obtain  the same result   
by  suitably    deforming
 the well  known DNLS soliton. Moreover, such 
 LF solitons  can also exhibit an unusual accelerating motion
  due to the time-dependent boundary condition of the deforming functions.

 Considering  
 the deformed DNLS hierarchy  we
can also  construct a novel  higher 
order deformation of the LF as well as the
DNLS equation.  

As a logical extension we  discover  the nonholonomic deformation for another
 form  of the 
DNLS equation (\ref {cll}),
proposed by Chen-Lie-Liu (CLL) \c{cll}, which is found  
 to share  the similar soliton dynamics and the mixed deformed integrable
hierarchy,  mentioned above.    

The arrangement of  the paper is  as follows. Sec. II presents the known
result of the undeformed  KN system. Sec. III
formulates  the nonholonomic deformation for the KN family, 
including its  
deformed two-fold integrable hierarchy. Sec. IV classifies these
  deformed  hierarchies,
 identifying   the LF  and the deformed DNLS equations.  Sec. V
 presents   the details on the LF equation  as a member of the 
  deformed KN hierarchy together with its higher deformations. Sec. VI
details similar result for the 
deformed  DNLS
equation. Sec. VII derives  in a simple way
 the exact  accelerating  soliton solution
of the deformed equations, including that of the LF equation. 
 Sec. VIII  introduces  the CLL equation,
 its nonholonomic   deformation and  the
related soliton solution with acceleration and other peculiarities.
 Sec. IX is the 
 concluding  section,
followed by the bibliography.

\vskip .8cm

\noindent {\bf II. KN SPECTRAL PROBLEM FOR THE DNLS  EQUATION}

For solving analytically the  DNLS equation  
  \be iq_t-q_{xx}+2i\epsilon  (| q|^2q)_x=0 , \ll{dnls}\ee
 through the inverse scattering method (ISM),
Kaup and Newell introduced a new type of
  spectral problem given by the Lax
pair \c{KN}: 
 \bea
U_{dnls}(\lambda)&=&-i \lambda^2\sigma_3
+\lambda U^{(0)},  \ \ U^{(0)}= q \sigma^++\epsilon q ^* \sigma^-, \ll{U} \\ 
  V_{dnls} (\lambda)& =& ( i\sigma_3U^{(0)}_x+  (U^{(0)})^{3})
\lambda-i \sigma_3(U^{(0)})^2 \lambda ^2
+2U^{(0)}\lambda^3-2i \sigma_3\lambda ^4 .  
 \ll{Vnn23}\eea
 The flatness condition
\be U_t-V_x+[U,V]=0 \ll{flatN}\ee
of this  Lax pair (\ref{U},\ref{Vnn23})  generates
the DNLS equation (\ref{dnls}) together 
with its complex conjugate. Note that  the time-Lax operator   (\ref{Vnn23}),
associated with   the second order dispersive
 DNLS equation,
contains positive powers    up to $ N_+  =4 $
in the spectral parameter $ \lambda $.
 The higher order  integrable equations in the 
 DNLS hierarchy
 can be obtained   similarly  from   a  Lax pair like   (\ref{flatN}) , where  the  space-Lax
operator $U(\la )$ remains the same, while 
 the time-Lax operator $V(\la )$ is  replaced by $V^{(N_+)}(\la )$, 
containing   
  higher  positive powers  in  $ \lambda $.
For example, for generating 
  a  $l$-th order dispersive
nonlinear equation  in the 
 DNLS hierarchy,   
 the 
$V^{(N_+)}(\la )$ matrix should contain 
 powers  in  $ \lambda $  up to  $N_+=2l $.

Recall that for extracting the 
   solutions of the DNLS equation through  the ISM
\c{KN}, the key role is played by the space-Lax operator (\ref{U}),
the  explicit knowledge of which 
 together with the required vanishing boundary condition for
the fields are sufficient for constructing the exact form of the  soliton
solution.
Only   at the final stage of the solution  we need to know the
asymptotic value of the time evolution 
operator (\ref{Vnn23}): $V_{dnls}(\la )|_{|x| \to
\infty}=V_\infty(\la )=-2i\lambda^4 \sigma^3 $, for determining the
time-dependence of the spectral data $S=\{
\rho(\lambda^2 ), \ 
\mbox{for}\  \lambda  \ \mbox { real} ;  b_k,
 \ \mbox{for}\  \lambda_k  \ \mbox { discrete}, \ k \in [1,N] \} $.
Pure soliton solution is obtained by setting the reflection coefficient
$\rho (\lambda^2 )=0, $ 
when the time evolution of the spectral data is obtained
 as $\dot
\lambda_k=0, \  \dot b_k=-4i \lambda^4_k b_k(t)  $ yielding 
$\lambda_k(t)=\lambda_k(0), \ b_k(t)=e^{-4i \lambda^4_kt}b_k(0) $, which 
  in turn  fix the soliton velocity $v_0 $ and the
frequency $\omega_0 $ of the wave enveloping the soliton.  
 
 Note that for all higher order
equations in the DNLS hierarchy the entire  picture of the ISM, 
including the structure of soliton solutions   basically remains the
same. Only the nontrivial
time-dependence of the spectral data,  determined  by
the boundary condition of the associated  time-Lax operator 
$V^{(N_+)}_{|x| \to
\infty}=-2i\lambda^{N_+} \sigma^3$   becomes  more general:
 $b_k(t)=e^{-4i \lambda^{N_+}_kt}b_k(0) $, changing
the soliton parameters  $v_0, \omega_0 $, accordingly.

Referring to \c{KN} for details on the N-soliton solution,
 we reproduce here the well
known 1-soliton  of the DNLS equation (\ref{dnls})
as
\be
q= 4\epsilon \eta \exp (2\theta-2i \sigma) \frac {\exp (4\theta)+ \exp
(-i \epsilon \gamma )} {(\exp (4\theta)+ \exp
(i \epsilon \gamma ))^2 }
\ll{1s} \ee
where \be \sigma=\xi x +2 \omega_0 t  , \ 
\theta=\eta (x -2  v_0t ) , \ 
\lambda_1^2= \xi+i\eta=\Delta ^2 e^{i \gamma}\ll{sth} \ee
with $\omega_0 $ as the frequency of the enveloping wave and $v_0 $  
as the constant soliton velocity. We shall see that the soliton solution of
the nonholonomic deformation of the DNLS hierarchy of equations can be
obtained in a easy way by deforming suitably the undeformed soliton
(\ref{1s}).
\vskip .8cm

\noindent {\bf III. NONHOLONOMIC DEFORMATION OF THE KN SYSTEM AND ITS 
MIXED INTEGRABLE  HIERARCHY} 



Recall that,   perturbation  as a rule destroys   the integrability of a
system. The  idea of  perturbation  through  
 nonholonomic deformation however is to perturb an
integrable 
system with a deforming function, such that under  suitable differential
constraints on the perturbing functions
 the integrability  of the whole system
is  preserved again. In a field theoretic models like the present one,
a constraint  given by differential relations 
(not  evolution equations) is equivalent to a nonholonomic
constraint.   For such novel  nonlinear systems, like the 
perturbed DNLS equation,  we can   derive 
all the important  features of a genuine integrable system, i.e. 
the  Lax pair,
 the integrable   hierarchy, 
the exact N-soliton solution etc. 

For implementing the concept of a nonholonomically 
 deformed  system we start not
from the perturbed equation, but  by
constructing the associated deformed
 Lax pair $U(\la ),V(\la ), $ the flatness condition of which  
generates then 
the required integrable perturbed equation.
 The space-Lax operator  for this perturbed system is   chosen to be the
 same as  that of the undeformed operator $U_0(\la ) $, while in the time-Lax
operator a deforming part     should be  added: $V(\la
)=V_0(\la )+ V_{def}(\la) $.
Therefore, since we keep  the spectral problem,   central
 to the  ISM,  unchanged the deformed systems  
automatically become exactly solvable by the ISM. However due to the 
 deformation of  the 
time-Lax operator, the time evolution of the  spectral data gets changed.
   As a result one gets
a deformed  soliton solution with
 changed dynamics, where interestingly, 
along with the basic field 
 the perturbing functions also take similar solitonic form.
A point of significant practical importance  in such perturbed
integrable equations is that, though  the ISM or other exact methods like
the RH method, the  dressing method, Hirota's bilinearization etc. are  applicable
here,
 one can  extract  in fact the same 
result    bypassing these
complicated  methods. For this one has to  take 
the   well known soliton solution of the
unperturbed  field  and change its dynamics by  simply    deforming
  the soliton  
 velocity and the frequency of the enveloping wave, which can induce
more complicated soliton dynamics, including 
 accelerated or decelerated soliton motion, as we will see below.  

\vskip .4cm

\noindent{\bf A. Nonholonomic deformation of the
   DNLS and the perturbed integrable
hierarchy}  

For generating the perturbed integrable systems through nonholonomic
deformation of the DNLS equation, we take the  
space-Lax operator  same   
  as in the unperturbed case
    (\ref {U}), while
  for  the time-Lax operator
  there could be 
different  choices. We may start from any    known integrable   hierarchy 
of  the DNLS 
 associated with $V_0(\la ) = V^{( N_+)}(\la ), $
which represents a polynomial in the spectral parameter $\lambda ^n $ with its
power $ n=0,1, \ldots $
 running upto an  arbitrary positive even integer $N_+ $ .  
    Such higher  operators can be constructed
systematically in the form 
 \be V^{(N_+)} (\la)= \sum_0^{N_+} (\la ) ^k
(\frac \partial {\partial x})^l(U^{(0)})^m, \  U^{(0)}=q\sigma^++\epsilon q^*
\sigma^-, \ \ N_+=\frac 1 2 (k+m)+l,  
\ll{Vn1}
\ee
where the summation indices  $k,l,m \geq 0$ are all positive integers  
   with
 the total  $1+N_+^2 $ number of terms appearing in this expression \c{JMP09}.
Therefore the corresponding  unperturbed system represents an integrable
 higher order equation in the +ve hierarchy of the DNLS equation 
with  $ \frac 1 2 {N_+}  $-th order dispersion.
Since we are considering the deformed system, the related
  time-Lax operator is  constructed as
 $V(\la)=V_0(\la )+ V_{def}(\la) $,  as stated above,
  with different  options for the deformed $  V_{def}(\la) $
operator containing  negative
powers of $ \lambda$  :
 \be V^{(N_-)}_{def}(\la)
= i  \sum_{j=0}^{ N_-}\lambda ^{-j}G^{(j)}.
\ll{VdefN}\ee
$G^{(j)}, \
j =0,1,2, \ldots N_-  $
is a set of  deforming matrices with
 arbitrary even integer $N_- $ , higher values of
which  indicate  higher orders of deformation. 
It is important to note that,
 while  the Lax
operators $U(\la )$ and the  undeformed  part $V_0(\la ) $ 
  involve only the basic field $q, q^\dagger $, the 
deforming part  $ V_{def}^{( N_-)}(\la)$ 
  contains only the perturbing  functions, expressed
  through  the matrix elements of the  deforming matrices  
  $G^{(j)}, \ j=0, 1,2, \ldots N_- $. The    
  nonholonomic   constraints on such perturbing functions are
 represented by  coupled differential equations,  defined in turn
by the flatness condition of the Lax pair for the deformed system.
With the above arrangement we  can build  the
 deformed  KN system 
again through the  condition (\ref {flatN}) as 
\be U_{0t}-V_{0x}+[U_0,V_0]=  V_{defx}-[U_0,V_{def}],\ll{dKN}\ee
yielding  at different powers of the 
parameter $\lambda $ the perturbed equation together with the   nonholonomic
constraints on the deforming matrix $V_{def} $.
 Note that though the basic field equation obtained
from (\ref{dKN}) may differ as 
  different members of the positive flow hierarchy
depending on the choice of $ N_+$ in  $V_0=V^{( N_+)}(\la ) $, the
perturbative part with       
  deforming matrices  $G^{(j)}, j=0,1,2, \ldots, N_- $ 
with the nonholonomic constraints would
 independently  yield  
another hierarchy of recursive
relations \bea
 G^{(0)}_x&=&[U^{(0)},G^{(1)}]-i[\sigma_3, G^{(2)}]
 ,
\nonumber \\
 G^{(1)}_x&=&[U^{(0)},G^{(2)}]-i[\sigma_3, G^{(3)}], \ \ldots \
 ,
\nonumber \\G^{(N_--2)}_x&=&i[U^{(0)},G^{(N_--1)}]-i[\sigma_3, G^{(N_-)}],
\nonumber \\ G^{(N_- -1)}_x&=&i[U^{(0)},G^{(N_-)}], \ \ G^{(N_-)}_x=0, 
 \ll{EEn} \eea
obtainable also from the flatness condition of the deformed Lax pair.
Remarkably, this deformed KN hierarchy  differs considerably
from  the deformed AKNS hierarchy found in \c{JMP09} due to
different spectral dependence between the corresponding Lax operators 
 $U(\la) $ and  the 
appearance  of an additional deforming matrix $G^{(0)} $ in the KN case, 
which leads also to a peculiar  $\ G^{(N_-)}(t) $  matrix
depending  only on the time variable  t,   as seen
from the last equation in (\ref{EEn}).

Using the explicit form of the Lax operators from (\ref{U}) and
(\ref{Vnn23}) or its higher  hierarchical
 form (\ref{Vn1}) together with the deforming operator
(\ref{VdefN}), 
one can  generate explicitly a two-fold  
 integrable hierarchy of the  perturbed KN-DNLS
family,
which we present below in detail.

\vskip .8cm

\noindent {\bf IV. CLASSIFICATION OF THE INTEGRABLE DEFORMATIONS IN THE KN
HIERARCHY } 


Before concentrating on the soliton solutions and detailed structure of the 
deformed DNLS 
and the LF
equations
 in the subsequent sections, we present here 
a classification for the   deformed KN hierarchy.
For this we  take the space-Lax operator  in the standard form
(\ref{U}), but select  a more general  time-Lax operator    
\be V^{(N_+,N_-)}(\lambda)=V^{(N_+)}(\lambda)+V^{(N_-)}_{def}(\lambda)
\ll{VN+-} \ee
with mixed flows in the spectral parameter 
exhibiting  a two-fold integrable  hierarchy. This    includes
a positive flow with \be
\ V^{(N_+)}(\lambda) = V_0+\lambda V_1+ \cdots + \lambda ^{N_+}
V_{N_+} , \ll{VN+} \ee
 representing the known higher order  integrable
 hierarchy of the DNLS equation, together with a negative flow:    
\be
 V^{(N_-)}_{def}(\lambda) =i( G^{(0)}+\lambda^{-1} G^{(1)}+ \cdots + \lambda
^{-N_-}
G^{(N_-)}), \ll{VN-}\ee
 describing  the deformation
hierarchy of the DNLS equation given through 
perturbation matrices  (\ref{VdefN}).
\vskip .4 cm

\noindent 
  {\bf A. Hierarchy without deformation}
  
This is generated by the pure positive flow with 
 arbitrary even integer $N_+ $  and 
  $N_-=0 $, having  the time-Lax operator:
$V^{(N_+,0)}=V^{(N_+)}$, as given in (\ref{VN+}).
 This case corresponds to   the  well known  KN-DNLS
 integrable hierarchy without
deformation \c{KN}, where each of the constituent matrices  $V_j $ in the
expansion $V^{(N_+)}=\sum_{j=0}^{N_+} \lambda ^{j}
V_{j} $  can be  build up systematically as (\ref{Vn1}) 
from  the known  building blocks
using only a dimensional argument, as shown in \c{JMP09}.
 The particular choice $N_+=4 $  yields the standard undeformed  DNLS
equation. 
\vskip .4 cm

\noindent 
 {\bf B. Pure deformation hierarchy }

 This is the complimentary case 
generated by the pure negative flow with  $N_+=0 $ and  $N_-$  as an 
 arbitrary even
integer with the time-Lax operator:
$V^{(0,N_-)}=V^{(N_-)}_{def}$, 
 given solely through deformation matrices  (\ref{VN-}). 
Such  deformations can be interpreted also as  novel integrable
perturbations of the lowest order equation 
$q_t=0 $  in the KN   hierarchy, where  
 we can generate  perturbed   
equations with different
 nonholonomic constraints for different    choices  of
 $N_-$ in (\ref{VdefN}).
We find remarkably that, 
any deformation with $ N_- < 2$ leads   to a trivial result with 
the   lowest order deformation   produced at $ \  N_-=2$ and
 all the subsequent  higher order deformations seem to be given by higher
even numbers   
$N_- \geq 2. $ 
We would analyze below in detail such     
deformations with concrete  examples.   
\vskip .4 cm

\noindent 
  {\bf C. General deformed case with
 mixed flow exhibiting a two-fold hierarchy}

 This mixed 
integrable hierarchy  is  
generated by both positive as well as negative flows with 
\ $N_+$  and $  N_-$ \ being 
 arbitrary even integers, 
 where the time-Lax operator is given in the general
 form (\ref{VN+-}): 
$V^{(N_+,N_-)}=V^{(N_+)}+V^{(N_-)}_{def} $, containing
  (\ref{VN+}) as well as  (\ref{VN-}).
The known integrable KN hierarchy of KN  is perturbed through 
another novel deformed  hierarchy,    
producing  thus    
    a two-fold hierarchy   of the deformed DNLS equations 
for  different   choices of $N_+,\ N_- $,
 few interesting cases of which we present below.
\vskip .4cm

\noindent 
i).  {\it Lenells-Fokas equation}: 

The first member with both positive and negative flow of the 
KN system is given by the lowest nontrivial 
 values  $N_+=2$ and $ N_-=2 $, with 
the time-Lax operator: $V^{(2,2)}=V^{(2)}+V^{(2)}_{def} $. Note that 
from the dimensional argument \c{JMP09} we must have $V^{(2)} \sim
U(\lambda)=\alpha U(\lambda) $ and $V^{(2)}_{def} $ should be constructed
from (\ref{VN-}) with minimum number of  deforming matrices $ G^{(j)}. j=0,1,2$.   
Surprisingly, one finds that
the recently proposed LF equation \c{LF}   is given in fact
by this  mixed  deformed KN hierarchy with the lowest order
deformation, where   $N_-> 2$ 
would generate a novel  integrable  hierarchy of 
the LF equation  with higher order deformations. 
We present the case of LF equation  in detail in the next section.
\vskip .4cm

\noindent 
ii).  {\it Deformed DNLS equation}

 The next entry in the mixed deformed
KN hierarchy is given by  
$N_+=4, N_-=2 $,  with the time-Lax operator:
 $V^{(4,2)}=V^{(4)}_{dnls}+V^{(2)}_{def}
$,
with $V^{(4)}_{dnls} $ as  the well known DNLS operator  (\ref{Vnn23}) and
 the deforming operator $V^{(2)}_{def} $, same as in  the above Lenells-Fokas
case
constructed from (\ref{VN-})  with $N_-=2 $.    
Therefore, since $N_+=4,\ N_-=0 $ corresponds to 
 the  unperturbed  DNLS equation, 
the present  case with  $N_+=4,\ N_-=2 $ would produce its 
 integrable perturbation  with minimal 
  nonholonomic constraint on the deforming
functions. Higher even integer  values of  $N_->2 $ would generate the
 DNLS hierarchy with higher
 order  integrable
deformations.
 
 We
 study below different cases of the deformed KN hierarchy   in detail.
 

\vskip .8cm

\noindent {\bf V. LENELLS-FOKAS EQUATION AND ITS DEFORMED HIERARCHY} 
 
Recently a new type of integrable 
NLS equation is  proposed   by  
 Lenells and Fokas and  solved it exactly for the  soliton solution through
the  RH method  \c{LF}.
 They also observed certain similarity of 
this  solution with the well known DNLS soliton \c{KN}. 
 We find intriguingly    that,  the LF equation 
 can be given by the minimal  nonholonomic deformation 
of the first  member in the  KN  hierarchy, where the subsequent 
perturbation can generate a  novel LF hierarchy with the higher  order
deformation.
 The link of the LF equation with the
first conservation law in  the negative flow  of the KN hierarchy 
has also been  identified  quite recently  by Lenells \c{lenells090}.
These findings    resolve thus  the mystery 
around the  integrability of the LF equation and  the resemblance of its 
solution with the DNLS soliton.
 As a bonus we can  get also the   explicit soliton solutions
of the LF equation in an easy way  by 
deforming  the well known DNLS soliton and  thus      
 can derive  the same result of  \c{LF,lenells09}
 through much simpler route, bypassing   
 the involved   RH   or  dressing method 
 employed  in  \c{LF,lenells09}.
Moreover the asymptotic form of our time-Lax operator given through the 
time-dependent boundary condition of the  
deforming functions can  lead
 to a remarkable possibility of creating  accelerating solitons
of the LF equation, 
similar to that found  for the deformed KdV 
equation \c{kundu082}. Following the scheme
described above  we can  construct along  with the LF equation a
novel integrable LF hierarchy  with increasingly higher   order deformation. 

 The associated Lax pair for the LF equation,  as classified above
for  the deformed  KN-DNLS hierarchy, 
  can be  constructed from the space-Lax
operator $U(\la) $ of the DNLS equation (\ref{U}) and the time-Lax
operator given through deformation of $V^{(2)}(\lambda)
=\alpha U(\lambda) $   as  
\bea
V^{(2,2)}(\lambda)&=&\alpha U(\lambda) +V^{(2)}_{def}(\lambda), \nonumber \\ 
V^{(2)}_{def}(\lambda) &=&
 i(G^{(0)}+ \lambda^{-1}G^{(1)}+ \lambda^{-2}G^{(2)}) \ll{vdef}\eea 
It is intriguing   that with a variable change 
$(x,t) \to (x, t-\frac 1 { \alpha}x) $ one can absorb the positive flow
part  $V^{(2)}(\lambda)=\alpha U(\lambda) $ from the time-Lax operator
(\ref{vdef}),
simplifying the LF equation and bringing this system to a pure deformation
 case discussed in Sec. IV B.
We can derive the constraint equations
on the deforming matrices $G^{(j)} , j=0, 1, 2$,
  from the general form (\ref{EEn})  for  this  minimal deformation 
with $N_-=2 $ as
\be
 G^{(0)}_x=[U^{(0)},G^{(1)}]-i[\sigma_3, G^{(2)}], \ 
 G^{(1)}_x=[U^{(0)},G^{(2)}], \ \ G^{(2)}_x=0,
 \ll{EE2} \ee
where $U^{(0)} =q\sigma^-+ \epsilon q^* \sigma^+$ as in (\ref{U}).     
Consistent with these relations   we may
  express the deforming matrices as
\be
 G^{(0)}=g(x,t) \sigma^3, \ G^{(1)}=
i(w(x,t)\sigma ^++ \epsilon w(x,t)^*\sigma ^-), \ 
G^{(2)}=a(t)  \sigma^3, \ll{Gs}\ee 
 through the deforming functions $g(x,t), w(x,t) $ and
 an arbitrary  function  
  $a(t) $, dependent only on time $t$.
We can  derive  the perturbed field equation from the relation (\ref{dKN})
by using the deformed Lax pair (\ref {U}, \ref{vdef}): 
\be U^{(0)}_{t}-\alpha U^{(0)}_{x}= -i[U^{(0)},G^{(0)} ] +i[\sigma
^3,G^{(1)}].\ll{dLF}\ee
Here the original equation is the LHS part, while the perturbation through
deforming matrices is given by the RHS terms. 
Using now the structure of $ U^{(0)} $ and  $G^{(0)}, G^{(1)} $ as in
(\ref{Gs}), we  obtain 
 the perturbed field equation from the matrix elements of (\ref{dLF})
 as  
\be
q_t-\alpha q_x+2w+2i gq=0
 \ll{qt}\ee 
 together with  its complex conjugate.  Note that here the nonlinear
 equation is generated by perturbation only, from an original linear equation
$q_t-\alpha q_x=0. $  
In this scheme $w $ and $g$ are like perturbing functions   which are
subjected to   the nonholonomic constraints obtained from (\ref{EE2}) as
\be
 g_x(x,t)=i \epsilon (qw^*-q^* w) , \ 
w_x(x,t)=- 2ia(t) q.
 \ll{gw}\ee 
Note that we can consider   equations (\ref {qt}) as a perturbed integrable
equation  with the nonholonomic constraints (\ref{gw}) imposed on the
perturbing functions. However since the original equation in this case is
only a  nonsignificant linear equation,
  an interesting alternative approach here would be  to obtain 
a new nonlinear integrable equation for the basic field by 
resolving the constraint relations and 
expressing all perturbating functions through this  field itself. Therefore   
for simplifying the set of equations (\ref{qt}-\ref{gw}) 
we   introduce a   potential field
 $q= u_x $, which  
solves  the constraints on the perturbing 
functions through the potential field as
\be
 g(x,t)=-2\epsilon a(t) |u |^2+c(t) , \ \ \mbox {and} \ \  
w(x,t)= -2ia(t) u ,
 \ll{gws}\ee 
  and 
 leads   (\ref{qt}) finally to  an evolution equation 
with a nonlinear derivative term in  the form
\be
 u_{xt} -\alpha u_{xx}+4ia(t) u+2ic(t)u_x-4i\epsilon  a(t) |u |^2u_x=0
 \ll{ut}\ee 
together with its complex conjugate and leaving no more constraint.
We note immediately that (\ref{ut}) is a nonautonomous, though integrable
generalization of the LF equation \c{LF} with 
arbitrary time-dependent
coefficients $a(t), c(t) $. For  constant choices of
 these functions: $a(t)=a_0, c(t)=c_0 $,
  on the other hand we get 
the corresponding autonomous LF equation and for a    particular 
choice  $a_0=1, c_0=0 $ we recover immediately from  (\ref{ut})   
 the recently proposed  LF equation
\be
 u_{xt} -\alpha u_{xx}+4i u+4i\epsilon   |u |^2u_x=0
. \ll{ut00}\ee 
It should be further noted that, a simple change in
time-variable: $t \to  t-\frac 1 { \alpha}x $, 
 as mentioned above,  can remove the dispersive
 term $-\alpha u_{xx} $   from the LF equation (\ref{ut00}) bringing it to 
a simpler form (\ref {u1t0}) considered below.
We show   in Sec. VII.
 that for the time-dependent $a(t)=a_0t $ and $ c(t)=c_0t $
 the more general 
 LF equation (\ref{ut}) allows 
novel accelerating soliton solution.
\vskip .4cm

\noindent{\bf A simpler  LF  equation}

A simpler equation equivalent to the  LF  equation,
 representing the DNLS hierarchy with pure 
deformation  can be derived from the general case at 
  $N_+=0, \  N_-=2$. 
The Associated Lax pair is given by a similar form (\ref{vdef}), with the term
 $V^{(2)}(\lambda)$ removed, i.e along with (\ref{U}) one should have 
$ \ 
V^{(0,2)}(\lambda)= V^{(2)}_{def}(\lambda) \ $, where the deformed Lax operator
 $V^{(2)}_{def}(\lambda) $,
is the same  (\ref{vdef})   as given for the LF case.  
We have mentioned already, that the LF equation can be brought to
this simpler case with the  time-Lax operator  $V^{(0,2)}(\lambda) $  by
 a simple change
in the time-variable.
As a result we get the set of equations given by  (\ref{qt}) without the 
$-\alpha q_x $ term:
\be
q_t+2w+2i gq=0,
 \ll{qt0}\ee 
together with its complex conjugate.
 The constraint relations (\ref{gw}) however remain the same,
leading to a unconstraint nonlinear integrable equation
\be
 u_{xt} - 4i(\epsilon  |u |^2u_x- u)=0
 , \ll{u1t0}\ee 
where we have taken $a(t)=1, c(t)=0 $ as for the LF equation. 
This simpler integrable equation (\ref{u1t0}),
 which is equivalent  to the LF equation 
under a  change of the time variable,
 is the least possible  pure deformation of the DNLS hierarchy. 
Soliton solutions of this  simplest member of the deformed 
KN-DNLS equations   can be found similarly by
deforming the known  DNLS soliton.
\vskip .4cm

\noindent{\bf B. Novel higher order
 deformation hierarchy  of the   LF  equation}

For generating the deformed hierarchy of the LF equation given by the higher
order nonholonomic constraint   on the perturbing functions, one has to 
take $N_- $  as an higher even integer, while keeping   $N_+ =2 $ fixed 
 as in the LF equation.
This would lead to the same perturbed equation   (\ref{dLF}) or equivalently
to (\ref{qt}), though 
due  to the increase in the number of  deforming matrices $G^{(j)} $  from 
$j=0,1,2 $ to $j=0,1,2, \ldots , N_- $ the constraint equations
(\ref{gw}) would be replaced by the more general deformed hierarchy 
(\ref{EEn}). Therefore it is no longer possible to solve the perturbing
functions through the potential field in a easy way like (\ref{gws}).
However it should be a challenge to derive some generalization of the LF
equation (\ref{ut}) for this higher deformed hierarchy  expressed again
through the potential field $u $.
 Therefore we take up the case of the first higher
deformation with $N_-=4 $, assuming $\epsilon =1 $  for simplicity. 
Here a larger set of  deforming matrices $G^{(j)}, \ j=0,1,2,3,4 $, 
 can be expressed as
\bea
 G^{(0)}&=&g(x,t) \sigma^3, \ G^{(1)}=i(w(x,t)\sigma ^++ w^*(x,t)\sigma ^-), \ 
G^{(2)}=a(x,t)  \sigma^3 \nonumber \\
G^{(3)}&=&i(b(x,t)\sigma ^++ b^*(x,t)\sigma ^-), \ G^{(4)}=c(t)  \sigma^3 
 \ll{Gs4}\eea 
 through the deforming functions $g(x,t), w(x,t), a(x,t), b(x,t) $ and
 an arbitrary time-dependent function  
  $c(t) $. Using the nonholonomic constraints (\ref{EEn}) for $N_-=4 $ and
introducing the potential field $q=u_x $ as above, we can partially remove
the constraints and get a novel higher order deformation of the LF equation
(\ref{ut00})  as
\be
 u_{xt} -\alpha u_{xx}+4i u+4i\epsilon   |u |^2u_x+2i\tilde g u_x+2\tilde w=0
, \ll{ut004}\ee
 where the constraints on the residual perturbing functions $\tilde g, \tilde
w $  are given by
\be
\tilde g_x=i( u_{x}\tilde w^*-u^*_{x}\tilde w) ,
 \ \tilde w_x=4c(t)(i|u|^2u_x-u) 
. \ll{wg4}\ee 
Note that the first part of the deformed LF equation (\ref{ut004})
is exactly same as the original LF  (\ref{ut00}), which however is  deformed   
now
by the {\it effective} perturbing functions, with  new 
 nonholonomic constraints (\ref{wg4}).   

We can construct the soliton solutions of the LF equation as well 
as all its higher deformations  using a simple trick of
 deforming the well known DNLS soliton, which we demonstrate in a later
section.


\vskip .8cm
 
\noindent{ VI. NONHOLONOMIC DEFORMATION OF
 THE DNLS EQUATION AND ITS DEFORMED HIERARCHY }

We concentrate now on a more interesting  
deformed DNLS equation,
 which may be obtained from  the general Lax pair (\ref{VN+-})
for the    particular choice  $N_+=4, \ N_-=2$.
Therefore in this case  
the  Lax operator $U(\la ) $ is the same as (\ref{U}), while   $V(\la )$ 
as mentioned already is given by 
\be
V^{(4,2)}(\lambda)=V^{(4)}_{dnls}(\lambda)+V^{(2)}_{def}(\lambda)  \ll{V12} \ee
where $V^{(4)}_{dnls}(\lambda)$
is the well known time-Lax operator
  (\ref{Vnn23}) of the  DNLS equation and  $V^{(2)}_{def}(\lambda) $ is  the
deforming part   (\ref{vdef}), same  as in the LF case.
The flatness condition of this  Lax pair yields the deformed equation
in the matrix form; 
\be  U^{(0)}_{t}- U^{(0)}_{x}- ((U^{(0)})^3)_x=i(-[U^{(0)},G^{(0)} ]
+ [\sigma
^3,G^{(1)}]),
\ll{dDNLSN}\ee
where the  matrix $U^{(0)} $
 is defined through the basic  fields $q,q^* $ as in 
(\ref{U}) and  the nonholonomic constraints
 on the perturbing matrices  are given
again by (\ref{EE2}).
In the scalar form  the perturbed   DNLS equation (\ref{dDNLSN}) may be
expressed as
\be
iq_t+ q_{xx}- 2i\epsilon (|q |^2q)_x=2( gq  -iw),
 \ll{qt1}\ee 
where the LHS is the original DNLS equation, while  the RHS  gives 
 its integrable perturbation through functions $w,g $ , with the nonholonomic
constraints on them defined as in (\ref{gw}).

  We may resolve these
constraints by introducing the   potential field $q=u_x $,
and
express the perturbing functions  through potential field as (\ref{gws}),
same as
in the LF case. Inserting this set of relations in    
the deformed DNLS (\ref{qt1}), we obtain a new   
integrable potential   DNLS  equation
in the form
\be
i u_{xt}+ u_{xxx}-2i  \epsilon  (|u_x |^2u_x)_x
-4a(t) u-2c(t)u_x+4\epsilon  a(t) |u |^2u_x=0.
 \ll{u1t}\ee 
This equation generalizes 
 the LF equation (\ref{ut}) and its simpler version (\ref{u1t})
 with the addition of another nonlinear derivative
term together with a higher dispersive term.
 Notice that considering 
$t \to it $ and  $u =iu_0 $ with    $u_0 $ as a    real
field,
 one can obtain from (\ref{u1t}) an integrable equation, which  
may   be interpreted  as a
novel  derivative generalization of the mKdV equation.

Deformed hierarchy of the DNLS equation 
may be constructed following the scheme  for the  LF system
 considered above, with the 
only  difference that, in the DNLS case one has to take 
$N_+ =4 $ as in (\ref{Vnn23}), while 
   $N_- $ for both these systems may be   any even
 integer as in 
(\ref{VN-}). This results  again to the 
same perturbed DNLS equation   (\ref{qt1}), though  the nonholonomic
constraints should be  given now  by a more general deformed hierarchy 
(\ref{EEn}). For $N_-=4 $, which corresponds to the first higher deformation 
similar to that considered above for the deformed LF equation, we obtain   
the same  perturbed DNLS equation (\ref{qt1}), though with   higher order 
nonholonomic constraints on the perturbing functions as
\be
 g_x=i(qw^*-q^*w), \ w_x=-2i(aq+b), \   
a_x=i(qb^*-q^*b), \
b_x=-2ic(t)q ,  
 \ll{Gsdnls4}\ee 
where the functions $g,w,a,b $ and $c(t) $ are
 the elements of the five deforming
matrices (\ref{Gs4}).  
 

\vskip .8cm

\noindent{\bf VII. EXACT ACCELERATING SOLITON SOLUTIONS OF THE  DEFORMED  DNLS
HIERARCHY }

As we have seen through the examples of the deformed DNLS  including
the LF equation,
for constructing  the nonholonomic deformations
 one has to  take the deformed time-Lax operator
$V(\la ) $ with an  additional  part $ V_{def}(\la )$, while
keeping   the same space-Lax operator $U(\la ) $ 
  \c{JMP09}.
 As a result  the x-dependent part in the deformed soliton solution 
remains the same as in the  undeformed case, while its  time-dependent part
 gets changed  by bringing in a new
{\it  time} in its dynamics, defined by the asymptotic 
value of $ \lim_{x \to \pm \infty} V(\la )
$. Interestingly, since the ISM for finding the  soliton solutions
 is
 based primarily on the spectral problem
$\Phi_x=U(\la) \Phi $ associated with the space-Lax operator,
 the structure of the N-soliton solution  remains the same even  for
 all the deformed DNLS equations in    their two-fold hierarchy with
arbitrary $
N_+, N_-$. Only at the final stage of the ISM we need to express the
time-dependent part
 of the N-soliton   by inserting the time-dependence of the
 related discrete spectral
 data: $\ {\lambda_j}, \  b_j(t), \  \ j = 1, 2, \ldots , N $,  
 determined through   
\be 
 \lim_{x \to \pm \infty} V(\la ) =i(2\lambda^{N_+} +\sum_{l=0}^{N_-}
\lambda^{-2l}c_{2l}(t)) \ \sigma^3, \ 
\ \mbox{as}\  \dot{\lambda_k}=0, \
 \dot  b_k(t)= i(2\lambda_k^{N_+}+\sum_{l=0}^{N_-}
\lambda_k^{-2l}c_{2l}(t))\  b_k(t). 
\ll{vinf} \ee 
 Here the time-dependent arbitrary
functions $c_{2l}(t), l=0,1,2, \cdots \ \frac {N_-} 2$
 are the boundary conditions of the deforming functions $
\lim_{x \to \pm \infty}
 G^{(j)}=ic_{j}(t)\sigma^3, \ j=0, 2, 4, \ldots   {N_-} $. 
Note that, in (\ref{vinf})  we  have restricted only to the 
terms with  even powers in $\lambda $, since 
it is evident from the    structure of the deforming matrices 
(\ref{Gs}) and (\ref{Gs4}) 
in the two   cases of deformation with $N_-=2$ and $N_-=4 $ that we
considered above,
 that only 
  the matrices $G^{j} $ with even  indices $j=0,2, 4 $ are diagonal 
and may have nontrivial 
 time-dependent asymptotic at the space-boundaries, while those with odd indices vanish
  at space infinities with  their
corresponding perturbing functions along  with the basic field.  
Note that the first term in (\ref{vinf}) comes from the 
standard hierarchy with positive flow $
V^{(N_+)}(\la )$, while the rest of the terms is due the  negative flow 
in this two-fold hierarchy, which 
describes the evolution along a new time.
 generated by (\ref {vinf}).    

Therefore we can  simplify
 the solution procedure enormously, since  instead of
applying  the ISM or any other 
sophisticated    method    for 
extracting the soliton
solution 
individually for every  equation belonging to  
this two-fold infinite hierarchy, we can simply take already available well known
N-soliton solution of the original DNLS equation  as in  Kaup-Newell \c{KN} and 
deform  the time-dependent part by  suitably choosing the values of $
N_{\pm} $ and thus  obtain  the N-soliton solution  for any 
 deformed equation in this
integrable  hierarchy almost without any effort.
For example, using the  identification of the Lenells-Fokas equation \c{LF}
as the $N_+=2, \ N_-=2 $ member in  the deformed DNLS hierarchy, we can reproduce
  its  soliton
solution quite easily,
 without going trough the  involved and lengthy 
procedure of the RH or the dressing method   used in  \c{LF,lenells09}. 
Moreover choosing the time-dependent boundary condition of the
deforming fields suitably we can find also a novel accelerating soliton in
such  deformed KN systems, observed already for the deformed AKNS  \c{JMP09}.

Based on our above argument we present the well known 1-soliton
 solution (\ref{1s}) of the
KN-DNLS  equation by deforming its parameters $\omega \to \omega(t), \ v \to
v(t)  $ 
\c{KN}
\be
q\equiv u_x= 4\epsilon \eta \exp (2\theta-2i \sigma) \frac {\exp (4\theta)+ \exp
(-i \epsilon \gamma )} {(\exp (4\theta)+ \exp
(i \epsilon \gamma ))^2 }
\ll{1sd} \ee
where \be \sigma=\xi x +2 \omega (t)  , \ 
\theta=\eta (x -2  v(t) ) ,
 \ \lambda_1^2= \xi+i\eta=\Delta ^2 e^{i \gamma}\ll{sthd} \ee
Such deformations can be extended easily to the N-soliton solution, which 
however we will not detail here. 
We stress that the 1-soliton solutions of all the equations in the  deformed DNLS hierarchy
can be expressed by the same expression
(\ref{1sd}), only the form of the {\it enveloping  frequency} $
 \omega (t)$ and  the {\it soliton velocity } $v(t) $ would be different 
for  particular equations,
depending on the  time-dependence of the corresponding  spectral data,
as obtained from the
general situation (\ref{vinf}).

For example in case of the standard DNLS \c{KN} with 
$ N_+=4, N_-=0$,  the time dependence
  obtained from  (\ref{vinf}) is
$ \lim_{x \to \pm \infty} V_{dnls}(\la ) =-2i\lambda^4\sigma^3 $, which gives
therefore  $  \omega (t) = \omega_0 t $ and 
$v (t) = v_0 t$, where constant parameters $\omega_0, v_0 $ are  given  in
as found through the ISM  in the pioneering work of  Kaup-Newell \c{KN}. 
 A similar  though  more generalized formulation, giving the
possibility of  accelerating solitonic motion,  will be presented below for 
 the  nonholonomic deformations of the DNLS hierarchy including the 
LF equation.

\vskip .4cm

\noindent {\bf A. Accelerating soliton  in the LF equation
as  deformation of the  DNLS soliton 
}

For the   LF equation   
(\ref{u1t}) the time dependence
  of the spectral data,  as explained above, is to be 
determined from (\ref{vinf})    with $N_+=2,N_-=2 $ as 
$b_1(t)= exp (-2i(\alpha\lambda_1^{2}t- \int dt ( c(t)+ \frac {a(t)} {
\lambda_1^2})))b_1(0). $
 Therefore following the above argument  we
can derive  the soliton solution for the LF equation 
from (\ref{1sd}) quite  easily by specifying the time-dependence of the 
  parameters with deformation 
as
\bea  \omega (t)= 2 Re ( \Phi (\lambda_1))=
- 2\alpha t \Delta ^2 \cos \gamma  +2  \int^t dt'(\frac {a(t)} {
\Delta^4} \cos  \gamma - c(t)),
  \nonumber \\  
v(t) = \frac 2 {\eta} Im (\Phi
(\lambda_1))= -2\alpha t -2\int^t dt' (\frac { c(t')} {
\Delta^2}   \sin  \gamma -\frac { a(t')} {
\Delta^4}) ,\nonumber \\
\mbox {where} \ \Phi(\lambda_1)= \alpha\lambda^2_1t-\int^t dt' (c(t')+ \frac
{a(t')} {
\lambda_1^2}).
\ll{0mvLF} \eea
Here the arbitrary time-dependent functions $c(t),a(t) $, 
which controls the dynamics  of the  soliton from the
space-boundaries are given by  the asymptotic
values of the deforming matrices $ G^{(0)}$ and  $ G^{(2)}$, respectively.  
 Thus we can obtain  the  soliton solution (\ref{1sd}) 
for   the
LF equation with the time-dependent soliton velocity $ v
(t)$ and frequency $\omega
(t)$ as (\ref{0mvLF}),    
 which can yield  accelerating solitons, generalizing
the  solution found in   \c{LF,lenells09}. In particular, for 
the choice of the functions $ c(t)=c_0t, \ a(t)=a_0t$ we obtain 
an  accelerating 
LF soliton:
\be
q\equiv u_x= 4\epsilon \eta \exp (2\theta-2i \sigma) \frac {\exp (4\theta)+ \exp
(-i \epsilon \gamma )} {(\exp (4\theta)+ \exp
(i \epsilon \gamma ))^2 }
\ll{1sLF} \ee
where \bea \sigma=\xi x  -\omega_0t+\omega_d t^2, \ 
& &  \mbox{where } \omega_0= 2\alpha  \Delta ^2 , \ \omega_d=  -((\frac {a_0 } {
\Delta^4}) \cos  \gamma +c_0)  , \nonumber \\ 
\theta=\eta (x  - (v_0t-v_d t^2)), \ & &  \mbox{where } v_0= 2\alpha , \ v_d=(\frac {c_0} {
\Delta^2}   \sin  \gamma -\frac {a_0} {
\Delta^4}).
\ll{sthLF} \eea
Integrating the   solution (\ref{1sLF}) we can get the
soliton solution of the LF equation (\ref {ut}) for the 
potential field  moving with a constant  acceleration as
\be
u=- 2i a(t)
 \sin \gamma  \frac {\exp (-(i \epsilon \gamma+2i \sigma))
} {\exp (2\theta)+ \exp
(i \epsilon \gamma -2\theta) }.
\ll{ulf}\ee
   
It is evident  that for $v_d(t) <0 $ we get an accelerating, while
for $v_d (t)>0 $ a decelerating   soliton.
Such a soliton for  the generalized LF  equation (\ref {ut})
 for the modulus of the potential   field
\be |u(x,t)|= \frac{2 \sin \gamma} {(
\cosh^2 2\theta - \sin^2 \frac{\gamma} 2)^{\frac 1 2}} ,\ll{1sumod}\ee
 given through  solution (\ref{ulf}) is  presented in
Fig. 1. 
\begin{figure}[!h]
\qquad \qquad
\includegraphics[width=6.5cm,height=5.4 cm]{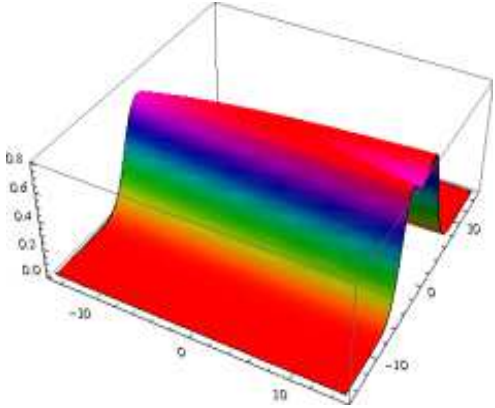}

\caption{ Dynamics of the   exact  soliton solution (\ref{ulf})
 for the potential field
$|u(x,t)| $   
of the generalized   Lenells-Fokas equation    (\ref{ut}). The unusual
 shape change
and bending of the soliton indicates its decelerating motion with 
time-dependent velocity $v_d(t)$.}
\end{figure}
Note that  (\ref{ulf}) gives a more general soliton 
solution of the LF equation, 
where for the  particular choice
 of the parameters with constant values: $c(t)=0  $ and
$a(t)=1, $  one 
 recovers exactly the constant velocity soliton found in \c{LF}.
 This
demonstrates that   we can extract 
 the same soliton solution of the LF
equation  by just deforming the parameters
of the
 known soliton \c{KN}  of the DNLS equation  and thus can 
avoid the  use of sophisticated  methods like the  RH
or the dressing method \c{LF,lenells09}. 

Another significant achievement
 is to construct in the same simple way the 
soliton
solutions for all higher deformed LF equations  by an easy extension of the 
time-dependence of the known    DNLS  or the LF soliton.
For example, for the first higher 
 deformation of the LF equation  
given by the higher nonlinear extension (\ref{ut004}), the soliton solution
can be expressed again in the form (\ref{ulf}), with similar
expressions for
 \[ \sigma=\xi x +2 \omega (t)  , \ 
\theta=\eta (x -2  v(t) ) \]
where  the time-dependence of the parameters should be  given now  in
a more extended form
\bea  \omega (t)= 2 Re ( \Phi (\lambda_1)), \ v(t) = \frac 2 {\eta} Im (\Phi
(\lambda_1)),
  \nonumber \\  
\mbox {where} \ \Phi
(\lambda_1)=\alpha\lambda^2_1 t - \int^t dt'(g_0(t')
+\frac {a_0(t')} {\lambda_1^2}+\frac {c(t')} {\lambda_1^4}),
%
\ll{0mvLF4} \eea
in comparison with the LF soliton (\ref{0mvLF}).
The arbitrary time dependent functions $g_0(t)$ and $ a_0(t)$  are
  asymptotic values of the 
perturbing functions $g(x,t)$ and $ a(x,t)$  as in (\ref{Gs4}), respectively  
and   can naturally  be taken also as constant
parameters.

Similarly, the N-soliton solution of the LF equation can  be 
obtained avoiding the involved methods of \c{LF,lenells09},  by
deforming the known N-soliton of the undeformed DNLS equation \c{KN} using (\ref{vinf})
with the choice $N_+=2,N_-=2 $, which  corresponds  to the LF equation.
We are however  not going to  reproduce here 
 these  otherwise simple steps.    

\vskip .4cm

\noindent {\bf B. Accelerating soliton  in the 
deformed   DNLS equation 
}

As we have stated,
 the soliton solutions for the integrable nonholonomic  deformation
of the 
 DNLS equation can be derived using general techniques like IST,
 Hirota's bilinearization etc. or by the RH and  the dressing method 
as done in \c{LF,lenells09}. However we find them here  in the
simplest way by deforming the well known DNLS solitons already obtained
in  \c{KN} through the ISM.  We   start therefore  
from the known undeformed soliton    (\ref {omvDNLS}) and deform
just its  parameters like frequency and the velocity using  
(\ref {0mvLF}) or (\ref {sthLF}),  as we done 
for extracting the  LF soliton.
Thus we derive the 
 1-soliton for the deformed DNLS equation 
(\ref{qt1}-\ref{gw}) or equivalently for  (\ref{u1t}), again  in the form
(\ref{1sLF})   yielding
\be 
|q|=\frac {2 \sqrt {2 }\eta }  {(\cos \theta+ 
\cos \gamma )^{\frac 1 2}}
\ll{1smod} \ee
 with the only difference that the undeformed
parameters like the  initial  frequency and the velocity
(\ref{sth}) are given now as in the   original DNLS soliton
  \c{KN}:
\be \omega_0 = 2 Re (\lambda^4_1)= 2(\xi^2-\eta ^2)=
2 \Delta ^4 \cos 2 \gamma,
 \ \  
v_0  = 2 Im (\lambda^4_1) /\eta= 4\xi =  4\Delta ^2 \cos  \gamma.
\ll{omvDNLS}\ee
The deformed DNLS soliton (\ref{1smod})  as
shown in Fig. 2a may  exhibit again 
 accelerated or decelerated motion,  similar to 
 the  LF case depicted in Fig. 1, 
but with a difference in their initial velocities $ v_0 $.
It is important to note here, that the
 perturbing functions $g, w $, which
 maintain the integrability of the
system are not any given functions as in  the usual perturbation theory, 
 but are determined consistently through internal dynamics and 
  can be expressed for the present models through
the potential field $u(x,t) $  as   (\ref{gws}).  Therefore  for the soliton
solution (\ref{ulf}) of the basic field these perturbing
 functions also    take  the 
  solitonic form, which 
  for the LF and the deformed  
DNLS equations  in particular  is as
\be
w(x,t)= 2ia(t) u 
\ll{1sw} \ee
expressed through  the solution (\ref{ulf})
  for the potential field $u(x,t) $
  and similarly using (\ref{1sumod}):
\be
 g(x,t)=2\epsilon a(t) |u |^2+c(t)
= c(t)+ 8\epsilon a(t) \frac{\sin^2 \gamma} {
\cosh^2 2\theta - \sin^2 \frac{\gamma} 2},
\ll{1sg} \ee
graphical representation of which is given in Fig. 2b.
Note that $w(x,t) $
 is a complex function with soliton solution (\ref{1sw}), while 
  $g(x,t) $  is a real function with  solution (\ref{1sg}), both
having time-dependent  velocity $v(t) $
 and  frequency $\omega (t) $  as in (\ref{sthLF}),
but additionally   with a
 time-dependent amplitude $a(t) $.
It should be  noticed  that,
 while at space infinities the perturbing function $w $
vanishes, $g $ goes to a time-dependent asymptotic $c(t) $,
and  drives the
solitons from the space-boundaries.
In this way the perturbing
functions,  themselves being  consistent soliton solutions,
  preserve the 
integrability of the system and can  force the   
basic field soliton to accelerate or decelerate.
Such perturbations thus can play potentially 
important role in practical
applications as we discuss later. 
\begin{figure}[!h]
\qquad \qquad
\includegraphics[width=5.5cm,height=4.4 cm]{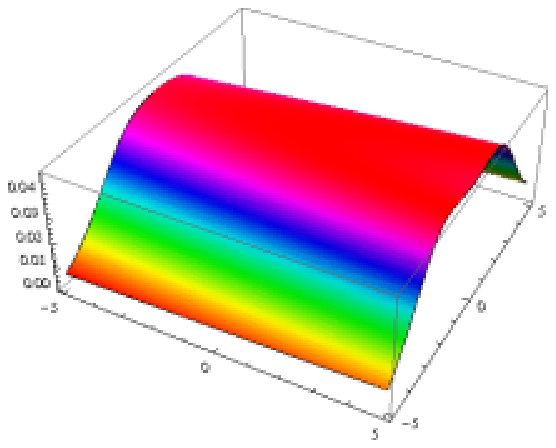}
 \qquad \qquad \ \ \
\includegraphics[width=5.5cm,height=4.4 cm]{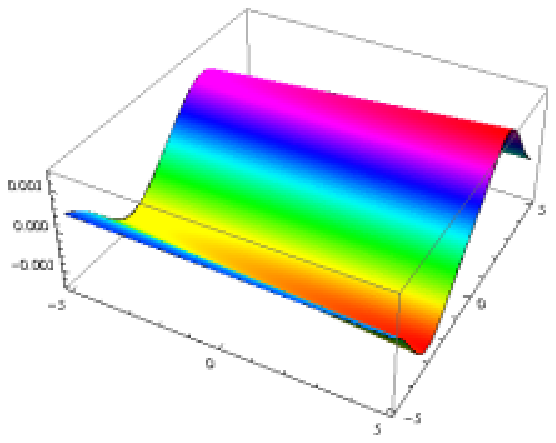}

  \hskip 4.2 cm  a) \hskip 6.8cm  b) 
\caption{ Exact  accelerating solitons  in the nonholonomic 
deformation of the  DNLS equation.
a) Soliton for the basic field $|q | $ corresponding to  
 (\ref{1sLF})  with (\ref{omvDNLS}).
 b) Consistent solitonic form (\ref{1sg}) for the perturbing field $ g(x,t)$  
 expressed  through the potential field
$|u(x,t)|^2 $. Notice the nontrivial boundary value of the perturbed
function.}  
\end{figure}

The higher N-soliton of the nonholonomic deformation of the  DNLS equation can
 similarly  be derived by
deforming the known N-soliton of the standard  DNLS equation \c{KN}. Likewise  the
solutions of all integrable equations in the deformed hierarchy 
 can  be obtained from  those of  the original 
undeformed integrable hierarchy by suitable   deformation of the solitonic
parameters as described above. In such deformed hierarchies, for every 
  fixed $N_+$, one may choose $N_- $  as   any  
even integer values,  like $ N_-=2,4$, as we have considered above, 
generating thus a novel two-fold  hierarchy.


\vskip .8cm

\noindent{\bf VIII. NONHOLONOMIC DEFORMATION OF THE CLL EQUATIONS, ITS
TWO-FOLD HIERARCHY AND ACCELERATING SOLITONS
}

The CLL equation is another form of integrable   derivative NLS   equation
with a  different  type of nonlinearity \c{cll}. Following our approach
as applied above for the deformed DNLS equation, 
we  construct here the integrable nonholonomic deformation  for 
 the CLL equation.

\vskip .4cm

\noindent{\bf A. Integrable CLL equation and soliton solution}

The   well known CLL equation given as
   \be i\psi_t-\psi_{xx}+i2\epsilon  | \psi|^2 \psi_x=0 , \ll{cll}\ee
 shares all  the  rich properties of an integrable
system, like
 commuting higher conserved quantities,
integrable hierarchy, Lax pair and exact soliton solutions. Moreover due to
its canonical Poisson bracket structure this  system allows
an exact quantum
 formulation with the algebraic Bethe ansatz solution \c{qDNLS}. 
It is important to note that the CLL equation  is   
 gauge related to the DNLS equation \c{jmp84} and therefore 
 it is 
intriguing  in the present context to find 
  the nonholonomic deformation of the CLL equation 
 and
examine  how the gauge  connection   affects its  
 relation
with the deformed DNLS derived  above.  

Since our approach for nonholonomic deformation starts with the Lax pair 
 we derive    them from those of the DNLS system  through a 
 gauge
transformation 
\be  U_{CLL}(\lambda)= h^{-1}U_{dnls}(\lambda)h-h^{-1}h_x, \  
V_{CLL}(\lambda)= h^{-1}V_{dnls}(\lambda)h-h^{-1}h_t. \ll{UVcll}\ee 
where  the gauge field is defined as 
\be h=e^{i \delta (x,t) \sigma^3}, \ \mbox{ with} \  \delta_x=\epsilon|q
|^2 =\epsilon|\psi|^2, \ \ \delta_t = i\epsilon(\psi^*\psi_x-\psi\psi^*_x)+
 6|\psi|^4. \ll{gt}
\ee  
Using the expressions  (\ref{U}) and
(\ref{Vnn23}) we can write down  explicitly the CLL Lax pair, the flatness
condition of which would generate the CLL equation (\ref{cll}). For example,
the
associated space-Lax operator, which is crucial for the ISM application, 
would take the form 
\be U_{CLL}(\lambda)=-i (\lambda^2+\epsilon|\psi|^2)\sigma_3
+\lambda ( \psi \sigma^++\epsilon\psi^* \sigma^-). \ll{U1}\ee
One can carry out the ISM independently  for extracting the soliton solution
for this equation, which however can be avoided by noticing a direct
relation between the CLL field $\psi $ and the DNLS    
field $q$ given by the gauge connection as 
\be \ \psi=q e^{-i\delta (x,t) },  \ \  \delta (x,t)= \epsilon 
\int ^x|q|^2 dy = \epsilon \int^x|\psi|^2 dy  , \ll{psiq} \ee
 which constructs the CLL soliton
 from 
(\ref{1s}) and (\ref{1smod}) as
\be \psi= 4\epsilon \eta  \frac{ \exp (2\theta-2i \sigma)}{\exp (4\theta)
+ \exp(i \epsilon \gamma ) },\ll{1scll} \ee
with the same expressions (\ref{sth}) for $ \theta, \sigma$ and 
(\ref{omvDNLS}) for   the unperturbed parameters  $\omega_0 $ and  $v_0 $. The
soliton solution under nonholonomic deformation, as we see below, can be
obtained by deforming the solitonic parameters (\ref{1scll})  in the
same way as done above.

\vskip .4cm

\noindent{\bf B. Nonholonomic deformation of the CLL equation and
accelerated soliton solution}

For constructing the deformed CLL equation  we follow  the above procedure by
taking the space-Lax operator $U_{CLL}(\lambda )$  as (\ref{U1}), while adding a deforming part 
(\ref{vdef}) to the time-Lax operator $V_{CLL}(\lambda)$.
The deforming matrices $G^{(j)}, \ j=0,1,2 $
  may be taken in the same form (\ref{Gs}), 
though   due the gauge 
transformation  they
 are linked in this case to  (\ref{Gs}) 
as $ G^{(j)} \to h^{-1}G^{(j)}h $.
Using therefore the deformed Lax operators we can derive
 the integrable perturbed CLL
equation  as 
\be i\psi_t-\psi_{xx}+2i\epsilon  | \psi|^2 \psi_x=2( g\psi-iw),
\ll{clld}\ee
and its complex conjugate, together with  
 a different set of  nonholonomic constraints on the perturbing functions:
\be
 g_x(x,t)=i { \epsilon } (\psi w^*-\psi^* w) , \ 
w_x(x,t)= 2ia(t) \psi-i \epsilon |\psi |^2w,
 \ll{gw1}\ee 
where $w $  in fact 
is gauge related to  (\ref{gw}) through a transformation like
(\ref{psiq}).
It is crucial  to notice here  that,   in spite of the 
gauge equivalence between  the  unperturbed  DNLS and the  CLL  equations,
their respective nonholonomic  deformations   
 exhibit
different structures as evident from (\ref{gw1}). 
As a result 
the resolution of  the constraints (\ref{gw1})  by
introducing  the   potential field,
 as done   for the deformed DNLS, 
clearly fails here, since in $w_x $ as seen from (\ref{gw1})  an
extra   term appears due to the gauge transformation. Therfore unlike the
LF equation or the deformed DNLS discussed above, the perturbing  fields
can not be eliminated   from the deformed set of the CLL equations
(\ref{clld})-(\ref{gw1}) through the local potential field and one 
 has to consider (\ref{clld}) as the 
perturbed CLL equation with nonholonomic constraints (\ref{gw1}).
 The whole perturbed system however remains
 integrable with
the associated Lax pair defined as above.
 The soliton solution for  this
deformed CLL equation  takes again the form  (\ref{1scll}), where
the expressions for $\theta, \sigma $ should be replaced  as (\ref{sthLF}). 
Note that
the unperturbed parameters are given here  as (\ref{omvDNLS}), while 
  the time-dependent
deformed frequency $\omega_d (t)$ and the   soliton
velocity $v_d(t) $ are  in the form (\ref{sthLF}). This would
result again to an accelerating soliton, where the
 solution $ |\psi|$ is  the same  as that for the
deformed DNLS case $|q|$, shown in Fig. 2a.    
It is important to note here that 
in the above constraint relations   (\ref{gw1}),
we have considered the same  solution  (\ref{Gs}), which  however
  is the  simplest  possible solution in the present case.
The deforming
matrix $G^{(2)} $   in fact can have a  more general structure   with
 nontrivial off-diagonal elements expressed through the function
$e^{\delta (x,t)} $ defined in (\ref{psiq}), since the last constraint
 in (\ref{EE2}) is  extended now  to a more general form:
$G^{(2)}_x+i \epsilon |\psi |^2 [\sigma^3,G^{(2)}] =0$.
  This   situation  
would naturally make the constraint (\ref{gw1}) more complicated,
 yielding another novel
integrable perturbation of the CLL equation.

Exploiting the integrability of the deformed CLL equation
 we can construct
its two-fold integrable hierarchy including the  
higher  deformations, following the same recipe as above for the deformed
 DNLS equations. 
For this we can use again the same space-Lax operator (\ref{U1})
 and construct   the 
generalized  time-Lax operator (\ref{VN+}, \ref{VN-}) from those of the
corresponding deformed DNLS by using the known gauge transformation
(\ref{UVcll},\ref{gt}). Note however  that   the higher order 
 nonholonomic constraints on the deforming matrices $G^{(j)}, \
j=0,1,2,\ldots $  are no longer same  as  (\ref{EEn}), though
 may be derived  from it through  the gauge
transformation (\ref{gt}), inducing the changes like
\[G^{(j)} \to h \tilde G^{(j)}h^{-1}, \ \ G^{(j)}_x \to  h (\tilde
G^{(j)}_x +[h^{-1}h_x,  \tilde G^{(j)}])h^{-1}. \]
We however will not reproduce here such hierarchical equations explicitly.
\vskip .8cm

\noindent{\bf IX. CONCLUDING REMARKS}

 The integrable  
nonholonomic deformation  of  the  Kaup-Newell 
class including the
 derivative 
NLS   equation, which  remained  as a challenging unsolved problem has been
solved here completely with the construction of a two-fold integrable
hierarchy with higher order deformations.
As an important application we identify the recent
 Lenells-Fokas  equation as a member of this deformed DNLS hierarchy, which 
allows us to bypass the involved procedures of
  \c{LF,lenells09} and find its soliton 
solution  as a simple deformation of the
known DNLS soliton. The soliton solutions in such deformed systems are
shown to exhibit unusual properties like
 accelerating (decelerating) solitonic motion, induced by the
 perturbing functions, which  control such motions 
from the space-boundaries.

The nonholonomic deformation of the Chen-Lee-Liu derivative
NLS equation though can be  constructed similarly,   
  gauge relation between the  undeformed CLL  
 with the Kaup-Newell DNLS equation, is not extended straightforwardly 
to their corresponding deformed equations. 
Nevertheless the soliton solution of the deformed CLL equation together with 
that for  its deformed   hierarchy can be constructed in a similar way.

A promising  potential  application of the  nonholonomic
deformation for  the family of DNLS equations arises when such deformations 
are interpreted as perturbations subjected to differential constraints,
 which
keep the whole system   integrable. 
Such perturbed DNLS equation can be considered 
 therefore as a novel coupled system of 
the DNLS and a Maxwell-Bloch type equation and may have important
applications in the fiber optics communication boosted by a doped laser
system, extending the idea of \c{NLS-MB} to a new direction \c{kundu10}.

 \end{document}